\documentclass[twoside,a4paper]{article}
\usepackage{fleqn,epsf,graphicx}
\usepackage{authblk}
\usepackage[utf8]{inputenc} 
\usepackage{fancyhdr}
\usepackage{indentfirst}

\title{PALS investigations of free volumes thermal expansion of J-PET plastic scintillator synthesized in polystyrene matrix}

\author{A.~Wieczorek$^{1,}$ $^{2}$, B. Zgardzińska$^{3}$, B. Jasińska$^{3}$, M. Gorgol$^{3}$, T.~Bednarski$^{1}$, P.~Białas$^{1}$, E.~Czerwiński$^{1}$, A.~Gajos$^{1}$, D.~Kamińska$^{1}$, Ł.~Kapłon$^{1,}$ $^{2}$, A.~Kochanowski$^{4}$, G.~Korcyl$^{1}$, P.~Kowalski$^{5}$, T.~Kozik$^{1}$, W.~Krzemień$^{1}$, E.~Kubicz$^{1}$, S.~Niedźwiecki$^{1}$,M.~Pałka$^{1}$, L.~Raczyński$^{5}$, Z.~Rudy$^{1}$, O.~Rundel$^{1}$, N.G.~Sharma$^{1}$, M.~Silarski$^{1}$, A.~Słomski$^{1}$, A.~Strzelecki$^{1}$, W.~Wiślicki$^{5}$, M.~Zieliński$^{1}$, P.~Moskal$^{1}$}

\begin{document}

\maketitle

	$^{1}$ Faculty of Physics, Astronomy and Applied Computer Science, Jagiellonian University, 
S. Łojasiewicza 11, 30-348 Kraków, Poland

       $^{2}$ Institute of Metallurgy and Materials Science of Polish Academy of Sciences, 
W. Reymonta 25, 30-059 Kraków, Poland
       
             $^{3}$ Department of Nuclear Methods, Institute of Physics, Maria Curie-Sklodowska University, Pl. M. Curie-Sklodowskiej 1, 20-031 Lublin, Poland
       
       $^{4}$ Faculty of Chemistry, Department of Chemical Technology, Jagiellonian University, 
R. Ingardena 3, 30-059 Kraków, Poland
       
            $^{5}$ Świerk Computing Centre, National Centre for Nuclear Research, 
             A. Soltana 7, 05-400 Otwock-Świerk, Poland
             
      $^{6}$ High Energy Physics Division, National Centre for Nuclear Research,
A.~ Soltana 7, 05-400 Otwock-Świerk, Poland\\

%{PACS: Put your PACS codes here}

\begin{abstract}
  The polystyrene dopped with 2,5-diphenyloxazole as a primary fluor and 2-(4-styrylphenyl)benzoxazole as a wavelength shifter, prepared as a plastic scintillator was investigated using positronium probe in wide range of temperatures from 123 to 423 K. Three structural transitions at 260 K, 283 K and 370 K were found in the material. In the o-Ps intensity dependence on temperature, the significant hysteresis is observed. Heated to 370 K, the material exhibits the o-Ps intensity variations in time. 
\end{abstract}

%
%  The body of the paper starts here
%

\section{Introduction}

Positron Annihilation Lifetime Spectroscopy (PALS) is commonly used technique in the investigations of the free volumes of polymers and their thermal properties from about three decades {\cite{a2}}. The main reason is that in the vast majority of known polymers a positronium formation and trapping is observed. Positronium mean lifetime value can be applied to estimate the sizes of free volumes present in the structure. The idea proposed by Tao {\cite{a15}} and Eldrup et al. \cite{a3} assumes free volume in the matter as a spherical potential well (infinite in depth) in which positronium annihilates. It enables to find the relationship between mean 
o-Ps lifetime value in the trap ($ \tau_{po} $) and the radius of the free volume (R):

\begin{equation}
\tau _{po}=0.5\cdot \left ( 1-\frac{R}{R+\Delta R}+\frac{1}{2\pi }sin\frac{2\pi R}{R+\Delta R} \right )^{-1} 
\end{equation}

where $ \Delta $R=0.166 nm is an empirical parameter describing the overlap of Ps wave function with surroundings. It was discussed that free volumes in many materials are not spherical and modification of proposed model is needed \cite{a5,a6}. However in the case of amorphous polymer, where shapes and sizes of the free volumes are not precisely defined and free volumes are interconnected, it is enforced to apply Tao-Eldrup formula in its original version.
	PALS technique was successfully applied to investigate many properties of polymers like free volume thermal expansion, structural transition, free volume fraction, number of free volumes or effects of irradiation in presence of radioactive source \cite{a2,a16,a19}. 
Plastic scintillators are commonly used in many detectors of radiation and for this reason they are continuously under development. For example, they are the key part of positron emission tomography device which is being constructed by the J-PET collaboration \cite{a9,a13,a14}. In this paper scintillator (hereafter referred to as J-PET scintillator) prepared with polystyrene matrix doped with 2,5-diphenyloxazole and 2-(4-styrylphenyl)benzoxazole (Table 1) is investigated. The novelty of the concept of the scintillator preparation lies in the application of 2-(4-styrylphenyl)benzoxazole as a wavelength shifter (a substance shifting maximum of emission to particular wavelength range). This substance was used for the first time for this purpose \cite{a17,a18}. 
A pure polystyrene, because of the possible applications, was widely investigated using many techniques including PALS \cite{a8,a11,a12}. Previous papers indicate that ortho-positronium (o-Ps) intensity is unstable and depends on the thermal history of the sample. However, the material stability is a crucial element of its possible application as a scintillator in any device. Thus we decided to use the PALS technique to investigate the structural stability of J-PET scintillator with the use of PALS technique.

\section{Experimental}

\begin{large}
\textbf{Material} 
\end{large}
\vspace{0.5cm} 

Scintillator samples were prepared by dissolving proper amounts of scintillation additives in liquid monomer - styrene (Table 1). The bulk polymerization was thermally initiated and during 100-hour-lasting synthesis, optically homogeneous plastic scintillator samples were obtained and tested in further research.

\begin{table}[]
\centering
\caption{Constitution of the J-PET plastic scintillator.}
\label{my-label}
\begin{tabular}{|c|c|c|}
\hline
Base & Primary fluor & Wavelength shifter \\ \hline
\includegraphics[scale=0.1]{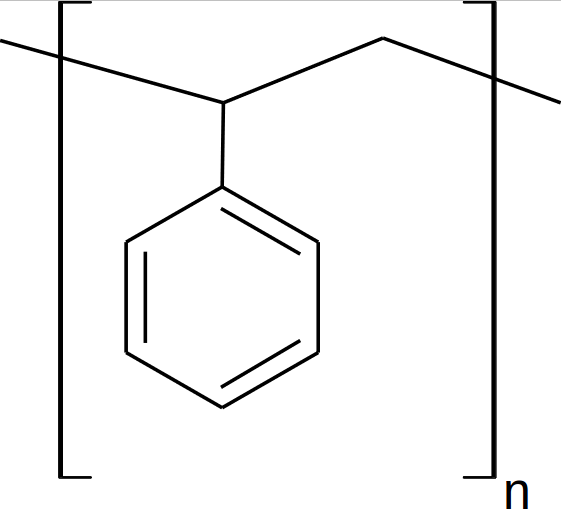}& \includegraphics[scale=0.1]{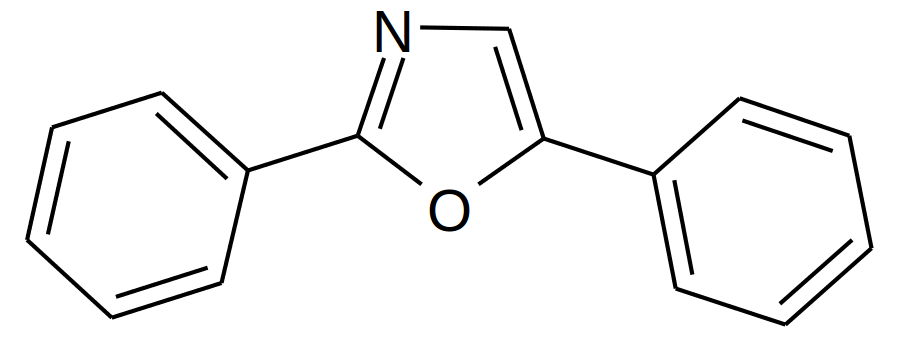}& \includegraphics[scale=0.1]{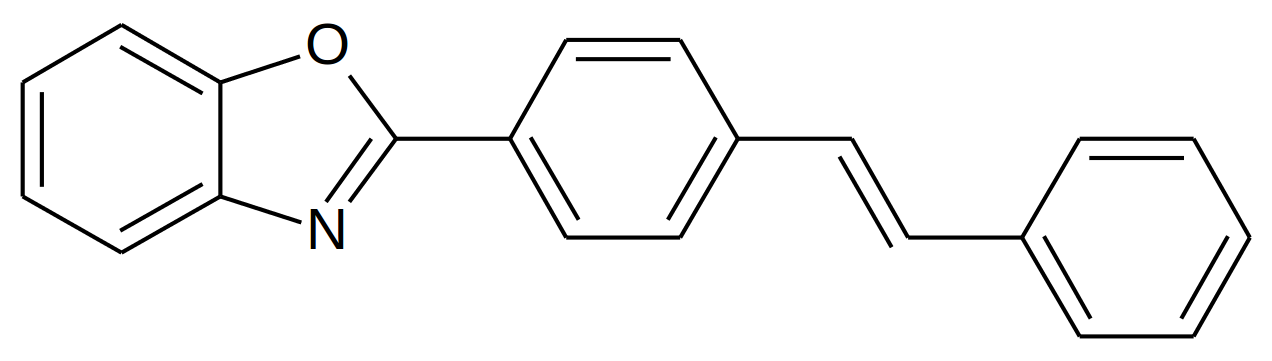} \\
Polystyrene (PS) & 2,5-diphenyloxazole (PPO) & 2-(4-styrylphenyl)benzoxazole\\ 

98\% wt. & 2\% wt. & 0,03\% wt. \\\hline

\end{tabular}
\end{table}

\section{Experimental technique}

	A standard "fast-slow" delayed coincidence spectrometer was used to carry out PAL measurements. The 22Na positron source of the activity of about 0.4 MBq, surrounded by two layers of sample, was placed inside a vacuum chamber. The $ \gamma $ quanta corresponding to positron creation inside the source and positron annihilation inside the sample were collected using two detectors equipped with cylindrical BaF2 scintillators of sizes $ \phi $1.5"x1.5" each. The resolution curve can be approximated with a single gaussian function of FWHM of about 210 ps. The PAL measurements were conducted within wide temperature range from 123 K to 423 K. A temperature increase was obtained using a resistance heater, while liquid nitrogen was used in order to cool the sample. Stabilization of temperature at various values was provided with a use of Shimaden FP21 PID controller.
A schematic diagram of temperature measurements sequence is presented in Fig.1. PAL spectra at each temperature, were collected for at least 2 hours and a total count number of 106 per spectrum was obtained.

\begin{figure}[h!]
\begin{center}
\includegraphics [scale=0.5] {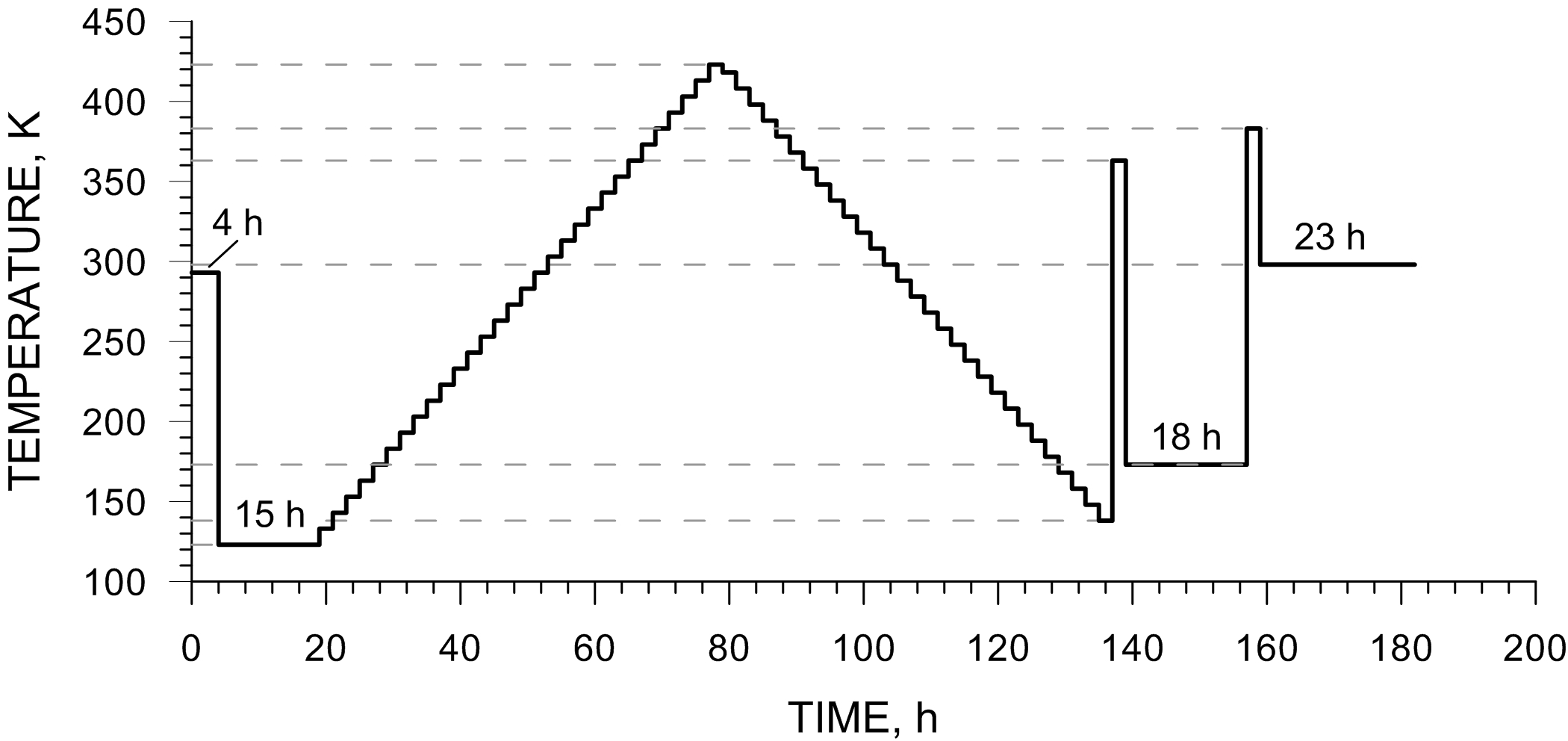}
\caption{A schematic temperature measurement schedule. Presented time values were highlighted only for those points, for which longer than 2 hour measurement were conducted.}
\end{center}
\end{figure}

The PAL spectra were analyzed with a use of LT 9.1 program \cite{a7}. Three discrete lifetime components of 170-190 ps, 380-570 ps and the one over 1.8 ns, corresponding to the annihilation of para-positronium, free annihilation and ortho-positrtonium, respectively, were found for each spectrum. 

\section{Results and discussion}

Positron lifetime spectra of J-PET scintillator sample were measured as a function of temperature and irradiation time. In Fig. 2 the changes of the o-Ps lifetime ($ \tau_{3} $), intensity (I3) and the lifetime of free annihilation ($ \tau_{2} $) as a function of temperature are shown. The intensity of o-Ps component is commonly accepted as the source of information about concentration and sizes of volumes in the medium via the o-Ps lifetime. We have noticed that the lifetime of free positrons e + varies with temperature, too. 

\begin{figure}[h!]
\begin{center}
\includegraphics [scale=0.46] {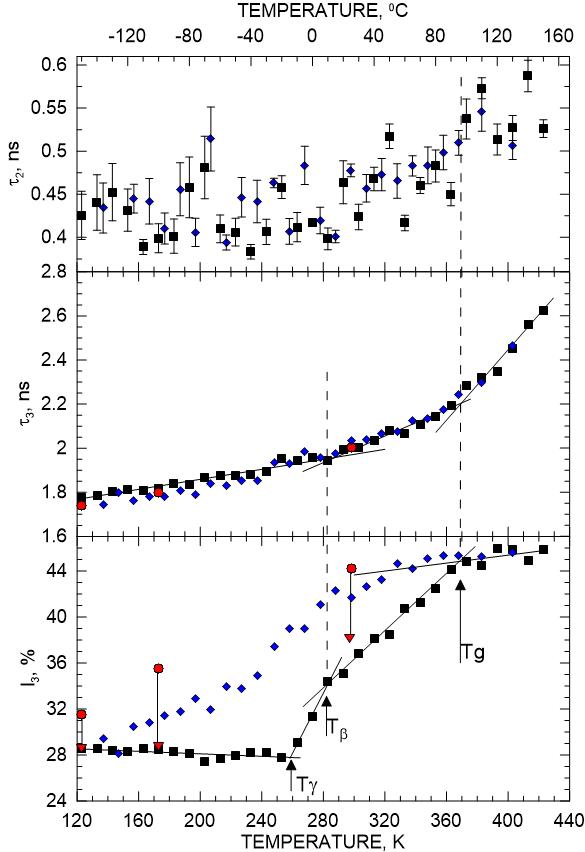}
\caption{Temperature dependence of lifetimes $ \tau_{2} $, $ \tau_{3} $, and intensity $ I_{3} $ in J-PET plastic scintillator. Points denote: squares - increasing temperature; diamonds - decreasing temperature; dots and triangles - first and last points in the measurements of irradiation effect (relaxation time described in details in Fig. 4). Dashed lines and arrows denote the transition temperatures.}
\end{center}
\end{figure} 

	The o-Ps lifetime increases linearly with temperature between 123 K and 260 K, starting from the value of about 1.75 ns to 1.95 ns and sustains at a constant value up to 283 K. Above this temperature the growth rate increases. Second change in the growth rate of lifetimes occurs at 370 K. For a pure polystyrene, a transition point at 87$^{\circ}C$ (360 K) was found previously \cite{a11}, but our result does not indicate it - the glass transition temperature is shifted of about 10 K. However the range of o-Ps lifetime values from our experiment agrees with mentioned paper. 
	The lifetime of free annihilation $ \tau_{2} $ as a function of temperature is also not constant. The changes in the slope of the dependence $ \tau_{2} $(T) at 283 K and at 370 K are noticed, but both, statistical scatter of points, as well as the measurements uncertainty does not allow to draw a clear conclusions except these that they conform the transition points observed in $ \tau_{3} $ dependence od temperature.
	The intensity $ I_{3} $ remains constant (~28\%) up to 260 K ($ T_{\gamma} $). Above this temperature it drastically rises, and at 370 K reaches a value of about 45\% (Tg). The growth rate collapses at 283 K ($ T_{\beta} $), the change in the o-Ps lifetime values was also observed at this temperature. This coincidence between the “smooth” changes in lifetime and “stepwise” changes in intensities confirms our conclusion that this effect is connected to structural transition, in contrary to undopped polystyrene. Thus in the investigated material three phases exist.
	It was stated that pure polystyrene can exist in three crystalline phases, as well as in the amorphous phase, depending on the way of sample preparation \cite{a1}. We expect that the studied sample has an irregular structure (amorphous) usual for atactic polystyrene but the presence of additional structural transitions ($ T_{\gamma} $, $ T_{\beta} $) suggests that the applied additives during material synthesis could modify the structure. Differences observed in two PAL measurements (for pure polymer, described in ref \cite{a11} and J-PET scintillator) are possible due to the doping of the polymer; even small impurity may significantly affect the organization of molecules and it can also shift the glass transition point (Tg). That supposition seems to be confirmed by the results of investigation done in polypropylene copolymers and blends \cite{a8}, where additional structural transition is observed. The existence of additional structural transitions may also result from the formation of small crystalline regions in the amorphous structure. One can find in the literature \cite{a1} information about transitions in polystyrene at low temperatures(between 283 K and 333 K) which would be identified with the structural transition point found in our results in J-PET polystyrene (at 283 K). Similar transition was found in another polymers like poly($ \varepsilon $-caprolactone) \cite{a2} and poly(methyl methacrylate) \cite{a16} too. Analysis of the o-Ps lifetime as a function of decreasing temperature and as a function of time shown complete repeatability and reproducibility of first results. Structural transitions ($ T_{\gamma} $) visible clearly in line course of $ I_{3} $ are almost invisible in  $ \tau_{3} $. It indicates that this transitions refer to energetic changes within molecule, not geometrical reconfiguration within a group of molecules. 
	In general the o-Ps intensity depends on many factors: material purity, thermal history, the direction of temperature change, rate of temperature change, the time of $ \beta+$ irradiation, etc. These factors modify the intensity, and therefore interpretation of results requires a series of additional tests. In the paper we focus on the selected aspects only. In Fig. 2 the PALS parameters changes as a function of increasing (squares) and decreasing (diamonds) temperature are shown. 
The experiments performed as a function of the decreasing temperature shown a significant hysteresis – intensities arrangement occurs below 140 K. The measurements as a function of increasing and decreasing temperature were repeated for the same sample. PAL spectra parameters were reproduced with an accuracy of statistical dispersions at the same rate of change in temperature. $ I_{3} $ depends significantly on the thermal history.
	Because the difference in $ I_{3} $(T) with the direction of temperature changes (thermal history), the additional measurements of stability over time were made. The PAL spectra (without any thermal history) was collected at room temperature (298 K) and the $ I_{3} $ is relatively high ~ 44\% (in \cite{a12} no variations of $ I_{3} $ is observed, it simply swings in the range 27-33\%). Next measurements were performed as a function of temperature according to the schedule presented in Fig 1. The sample was cooled to 123 K and stored at this temperature for 15 h. During that time values of  $ \tau_{2} $ and  $ \tau_{3} $ were stable, but the intensity $ I_{3} $ rapidly decreases (Fig.3). After this time the series of measurements vs. temperature is made, as shown Fig. 1. Then the sample was heated close to Tg, to eliminate the influence of a long time of irradiation (~140 h) \cite{a11,a12}, and the $ I_{3} $ stability at selected temperatures was investigated. 
	The temperature jumping with rate 5 K/min was performed from 363 K and 383 K to 173 K and 298 K, respectively (see Fig.1). As shown in Fig. 3, the o-Ps intensity is unstable in time wherein the time constants are shorter with lower temperature. 

\begin{figure}[h!]
\begin{center}
\includegraphics [scale=0.34] {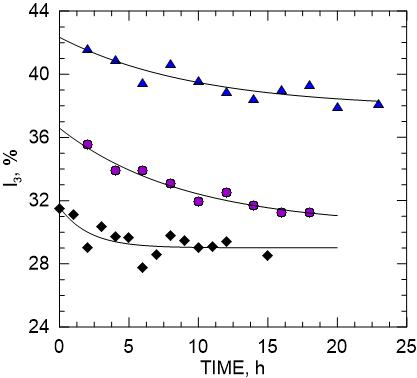}
\caption{Positron irradiation effect on the o-Ps intensity in J-PET scintillators at 123 K (diamonds), 173 K (dots) and 298 K (triangles)}
\end{center}
\end{figure}
	
	The time constant at 123 K is 2 h, at 173 K extends to 8 h, and at 298 K about 10 hours are needed to stabilize the intensity $ I_{3} $. Similar effects of time instability of $ I_{3} $ in modified polystyrene have already been observed \cite{a12}, and their time constant varies from about 3 to 30 h. Such measurements were performed at room temperature only and samples were not subjected to thermal treatment like in our case. Fast decrease of the intensity within few hours is ascribed to competition between the creation and decay of free radicals \cite{a12}. In our case the intensity changes are influenced additionally by the temperature distance between last point before jump and temperature of time measurement. The I3 dependence on time was also the subject of study in the paper by Peng et al. \cite{a11}. They have observed the increase of intensity value at low temperatures, however our material does not exhibit such tendency. Observed differences are caused probably by different methods of material synthesis.

	The size of spherical free volumes was calculated using the formula (1), and the results of these calculations are shown in Fig.4. Over 300 K the size of free volume is doubled.

\begin{figure}[h!]
\begin{center}
\includegraphics [scale=0.34] {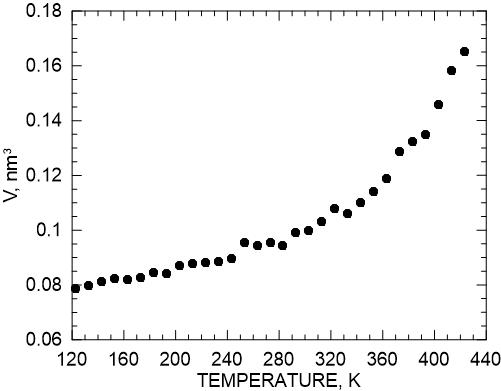}
\caption{The size of a spherical free volume as a function of temperature.}
\end{center}
\end{figure}

\section{Conclusions}

The polystyrene dopped with 2,5-diphenyloxazole and 2-(4-styrylphenyl)benzoxazole was investigated using PALS technique and three structural transitions at 260 K, 283 K and 370 K were observed (concluding from $ I_{3} $). The last transition point has been identified as the glass transition, which is also confirmed by the o-Ps and e+ lifetimes changes. In addition it was determined that in the studied material the o-Ps lifetime does not depend on the thermal history, yet the o-Ps production intensity reveals a significant hysteresis connected mainly to matrix material. However, it does not affect scintillation properties of the material. From our results it may be inferred that o-Ps intensity instability is not due to the structural instability of the investigated material. Structural transition $ \beta $ type ($ T_{\beta} $ in Fig.2.) may be connected to dopant percentage. However the final conclusions need additional studies of differental scannig calorimetry (DSC) or pressure - volume - temperature (PVT) measurements.  

\section{Acknowledgements}

We acknowledge technical and administrative support by T. Gucwa-Ryś, A. ~Heczko, M. Kajetanowicz, G. Konopka-Cupiał, W. Migdał and the financial support by the Polish National Center for Development and Research through grant INNOTECH - K1/1N1/64/159174/NCBR/12, the Foundation for Polish Science through MPD programme and the EU and MSHE Grant No. POIG.02.03.00 - 161 00-013/09.

%  The bibliography starts here
%  The first three items serve as the examples for
%  the citation of paper, book and article in proceedings.
%

\end{document}